\documentclass[a4paper,11pt]{article}
\usepackage{graphicx}
\usepackage{multirow}

\usepackage{color}
\usepackage{latexsym}
\usepackage{amssymb}
\usepackage{amsfonts}
\usepackage{amsmath}
\usepackage{indentfirst}

\newcommand{\cvd}{\hfill $\blacksquare$\bigskip}
\newtheorem{definition}{Definition}[section]

\newtheorem{proposition}{Proposition}[section]
\newcommand{\concc}[2]{\underset{#1}{\overset{#2}{\bigcirc}}}

\date{}
\author{Elena Barcucci\thanks{Dipartimento di Matematica e Informatica ``U. Dini'', Universit\`a degli
Studi di Firenze, Viale
 G.B. Morgagni 65, 50134 Firenze, Italy. {
 \tt \ elena.barcucci@unifi.it,\quad antonio.bernini@unifi.it,\quad stefano.bilotta@unifi.it,\quad
renzo.pinzani@unifi.it}}\and Antonio Bernini$^*$ \and Stefano Bilotta$^*$ \and Renzo Pinzani$^*$}

\title{Cross-bifix-free sets in two dimensions}

\begin{document}

\maketitle

\begin{abstract}
A \emph{bidimensional bifix} (in short \emph{bibifix}) of a square matrix $T$ is a square submatrix of $T$ which
occurs in the top-left and bottom-right corners of $T$. This allows us to extend the definition of bifix-free words and
cross-bifix-free set of words to bidimensional structures. In this paper we exhaustively generate all the
\emph{bibifix}-free square matrices and we construct a particular non-expandable \emph{cross-bibifix}-free set of
square matrices.  Moreover, we provide a Gray code for listing this set.
\end{abstract}


\maketitle

\section{Introduction}
A word $w$ over a given alphabet is said to be \emph{bifix-free} \cite{13} if and only if any prefix of $w$ is not
a suffix of $w$. A \emph{cross-bifix-free set} \cite{2} of bifix-free words (also called \emph{cross-bifix-free code}
\cite{7}) is a set where, given any two words over an alphabet any prefix of the first one is not a suffix of the second
one and vice-versa.

Cross-bifix-free sets, which are involved in the theory of codes and in formal language theory, are usually applied in
the study of frame synchronization which is an essential requirement in a digital communication systems to establish and
maintain a connection between a transmitter and a receiver. The problem of determining such sets is also related to
several other scientific applications, for instance in pattern matching \cite{8}, automata theory \cite{4} and pattern
avoidance theory \cite{5}. Several methods for constructing cross-bifix-free sets have been recently proposed as in
\cite{2,6,7}.

In this paper we introduce, probably for the first time, an extended version of the linear case in order to
generalize the topics concerning the cross-bifix-free sets of words to sets of matrices. Actually, within the formal
language theory, the extension to the bidimensional case of a concept is significant and interesting by itself. There
are several cases in the literature where a similar process is occurred. For example, in \cite{9} a
bidimensional variant of the string matching problem is considered for sets of matrices. Another interesting example is given by the
extension of classical finite automata for strings to the two-dimensional rational automata for pictures introduced in
\cite{1}. Moreover, it is worth to mention the problem of the pattern avoidance in matrices \cite{12}, which is a
typical topic in linear structures as permutations and words.

Since the theory of cross-bifix-free sets of word is widely used in several fields of applications, we are expected
that the extension to the two-dimensional case could have the same usefulness and it could constitute a starting
point for a fruitful and intriguing theory.

\bigskip

After a brief background and the needed definitions (Section 2), we start with the exhaustive generation of the
bibifix-free set of $n\times n$ square matrices for each $n\geq 1$ over a $q$-ary alphabet (Section 3), then we define
an its proper non expandable cross-bibifix-free subset (Section 4). Moreover, the particular structure of this set
allows us to obtain a Gray code for listing it in order to facilitate its possible utilities (Section 5). Finally, we
conclude with some hints for future developments (Section 6).

\section{Basic definitions and notation}
Let $\Sigma=\{0,1,\cdots,q-1\}$ be an alphabet of $q$ elements. A (finite) sequence of
elements in $\Sigma$ is called (finite) \emph{word} or \emph{string}. The set of all
strings over the alphabet $\Sigma$ is denoted by $\Sigma^*$ and $\Sigma^+=\Sigma^*\setminus \{\epsilon\}$, where
$\epsilon$ denotes the empty word. If $w\in\Sigma^+$ is a word, then $w^n$ is the word which consists of $n$ copies of
$w$.

Let $w=uzv$ a length $n$ string in $\Sigma^+$, then $u\in\Sigma^+$ is called a \emph{prefix} of $w$ and $v\in\Sigma^+$
is called a \emph{suffix} of $w$. A \emph{bifix} of $w$ is a subsequence of $w$ that is both its prefix and suffix. A
string $w \in \Sigma^+$ is said to be \emph{bifix-free} \cite{13} if and only if no prefix of $w$ is also a suffix of
$w$. We recall the following proposition which allows to check only the prefixes and the suffixes with length up to
$\lfloor \frac{n}{2}\rfloor$ in order to establish if $w$ is bifix-free (see \cite{13}).

\begin{proposition}\label{Nielsen}
A word $w=w[1]w[2]\ldots w[n]$ is a bifix-free word if and only if $w[1]w[2]\ldots w[i]\neq w[n-i+1]w[n-i+2]\ldots
w[n]$ for $i=1,2,\ldots,\lfloor \frac{n}{2}\rfloor$.
\end{proposition}

For example, the string $111010100$ of length $n=9$ over $\Sigma=\{0,1\}$ is bifix-free, while the string $100100100$
contains two bifixes, $100100$ and $100$.

\bigskip

Let $BF_n^q$ denote the set of all bifix-free strings of length $n$ over an alphabet of fixed size $q$. The following
formula for the cardinality of $BF_n^q$, denoted by $|BF_n^q|$, is well-known \cite{13}.

\begin{equation*}
\label{bifixfree} \left \{
\begin{array}{lll}
|BF_1^q|=q\\\\
|BF_{2n+1}^q|=q|BF_{2n}^q|\\\\
|BF_{2n}^q|=q|BF_{2n-1}^q|-|BF_{n}^q|
\end{array}
\right.
\end{equation*}

The related number sequences can be found in
the On-Line Encyclopedia of Integer Sequences:
A003000 ($q=2$), A019308 ($q=3$) and A019309 ($q=4$).

Given $q>1$ and $n\geq1$, two distinct strings $w,w'
\in BF_{n}^q$ are said to be \emph{cross-bifix-free} \cite{2} if
and only if no prefix of $w$ is also a suffix of
$w'$ and vice-versa.

For example, the binary strings $111010100$ and $110101010$ in $BF_{9}^2$ are
cross-bifix-free, while the binary strings $111001100$ and
$110011010$ in $BF_{9}^2$ have the cross-bifix $1100$.

A subset of $BF_{n}^q$ is said to be a \emph{cross-bifix-free set} if
and only if for each $w, w'$, with $w \neq
w'$, in this set, $w$ and $w'$ are
cross-bifix-free. This set is said to be \emph{non-expandable} on
$BF_{n}^q$ if and only if the set obtained by adding any other word in $BF_{n}^q$
is not a cross-bifix-free set. A non-expandable cross-bifix-free
set on $BF_{n}^q$ having maximal cardinality is called a
\emph{maximal cross-bifix-free set} on $BF_{n}^q$.

\medskip
In the following we give the notation we are going to use in the paper.
A two-dimensional (or bidimensional) string is a $n_1 \times n_2$ matrix
with entries from $\Sigma$.
In this paper we deal exclusively with the special case of square matrices, $n_1=n_2=n$.
An $n \times n$ square matrix $T$ will be sometimes represented by
$T[1 \cdots n, 1 \cdots n]$, when we need to point out its rows and columns. Fixed $r < n$, an $r \times r$ matrix $P$
is a
\emph{submatrix} of $T$, if the upper left corner of $P$ can be aligned with an element
$T[i,j]$, $1 \leq i,j \leq n-r+1$, and $P[1 \cdots r, 1 \cdots r]=T[i \cdots i+r-1,j \cdots j+r-1]$.
In this case, the submatrix $P$ is said to \emph{occur} at position $[i,j]$ of $T$.
A submatrix $P$ is said to be a \emph{bidimensional prefix} (in short \emph{biprefix})
of $T$ if $P$ occurs at position $[1,1]$ of $T$. Similarly, a submatrix $P$ is
a \emph{bisuffix} of $T$ if $P$ occurs at position $[n-r+1,n-r+1]$ of $T$.
A \emph{bidimentional bifix} (in short \emph{bibifix}) of a square matrix $T$
is a submatrix of $T$ which is both a biprefix and a bisuffix.

\begin{definition}
A square matrix $T$
is said to be \emph{bibifix-free} if and only if no biprefix of $T$ is also
a bisuffix of $T$.
\end{definition}
For example, considering $\Sigma=\{0,1\}$, the matrix
$T=\left(
  \begin{matrix}
    1&1&1& 1& 0\\
    1 &0 &1 &1& 1\\
    1 &0& 0& 1& 0\\
    0& 1& 1& 1& 0\\
    1& 0& 0& 0& 0
   \end{matrix}
\right)
$
is bibifix-free, while
$
M=\left(
\begin{matrix}
 1&1&1&1&0\\
 1&0&1&1&1\\
 1&0&0&1&0\\
 0&1&1&1&1\\
 1&0&0&1&0
\end{matrix}
\right)
$
is not bibifix-free, since a bibifix
$P=
\left(
\begin{matrix}
 1&1\\
 1&0
\end{matrix}
\right)$ of dimension $2\times 2$ occurs in $M$.

Analogously to the linear case, we have the following proposition which ensures that one must check only the biprefixes
and bisuffixes of dimension up to $\lfloor\frac{n}{2}\rfloor\times \lfloor\frac{n}{2}\rfloor$ in order to establish if
$T$ is bibifix-free.

\begin{proposition}
A square matrix $T[1\ldots n,1\ldots n]$ of dimension $n\times n$ is \emph{bibifix-free} if and only if
$P[1\ldots r,1\ldots r] \neq S[1\ldots r,1\ldots r]$,
$\forall r =1,2,\ldots,\lfloor \frac{n}{2}\rfloor$ where	
$P[1\ldots r,1\ldots r]$ and $S[1\ldots r,1\ldots r]$ are the
biprefixes and the bisuffixes of dimension $r\times r$ of $T[1\ldots n,1\ldots n]$.
\end{proposition}

\noindent
\emph{Proof.}\quad
If $T$ is bibifix-free, then the thesis follows directly from the definition of bibifix-free matrix.

\medskip

\noindent
Suppose $P[1\ldots r,1\ldots r] \neq S[1\ldots r,1\ldots r]$,
$\forall r =1,2,\ldots,\lfloor \frac{n}{2}\rfloor$. We have to check
that $P[1\ldots j,1\ldots j] \neq
S[1\ldots j,1\ldots j]$,
$\forall j> \lfloor \frac{n}{2}\rfloor$. We proceed ad absurdum.

Let $l\geq\lfloor \frac{n}{2}\rfloor +1$ such that
$P[1\ldots l,1\ldots l]=S[1\ldots l,1\ldots l]$. Then, their intersection, which is a square matrix of dimension
$(2l-n)\times (2l-n)$,  is a bibifix of $T$. If $2l-n\leq \lfloor \frac{n}{2}\rfloor$, the proof is completed since we have a contradiction.
Otherwise, we consider this bibifix which, read as bisuffix and biprefix, gives
rise to a new intersection. Such an intersection is again a bibifix of $T$. Then, by means of a recursive argument,
we finally obtain a bibifix of dimension less than $\lfloor \frac{n}{2}\rfloor\times \lfloor \frac{n}{2}\rfloor$, against
the hypothesis.

\cvd

In the next section we present the exhaustive generation of the bibifix-free set of $n \times n$ square matrices for each $n \geq 1$ over
a $q$-ary alphabet.

\section{Bibifix-free sets generation}

We indicate with $\mathcal{M}_n$ the set of all matrices $M[1\ldots n ,1\ldots n]$ with entries in $\Sigma=\{0,1,\cdots,q-1\}$ and we
denote by $\mathcal{P}^{\mathcal{M}_n}$ its power set (the set of its subsets).

\begin{definition}
Let $\varphi\colon \mathcal{M}_n\to
\mathcal{P}^{\mathcal{M}_{2n}}$ such that:

$$
\varphi(M)=
\left\{\left(
\begin{array}{c|c}
M[1\ldots n ,1\ldots n] &
\begin{matrix}
*&*&\ldots&*\\
*&*&\ldots&*\\
\hdotsfor{4}\\
*&*&\ldots&*\\
\end{matrix}

\\ \hline

\begin{matrix}
*&*&\ldots&*\\
*&*&\ldots&*\\
\hdotsfor{4}\\
*&*&\ldots&*\\
\end{matrix}

& M[1\ldots n ,1\ldots n]
\end{array}
\right): *\in \Sigma\right\}.
$$

\end{definition}

\noindent
If $M$ is a matrix of dimension $n\times n$, the operator $\varphi$ creates a set of matrices with
dimension $2n\times 2n$ where in each new matrix the two diagonal blocks of dimension $n\times n$ are equal to $M$
and the other entries are chosen from the alphabet. For example the matrix $\left(
\begin{matrix}
 \textbf{1}&\textbf{0}&0&0\\
 \textbf{1}&\textbf{0}&1&0\\
 1&1&\textbf{1}&\textbf{0}\\
 0&0&\textbf{1}&\textbf{0}
\end{matrix}
\right) \in \varphi\left(
\begin{matrix}
 \textbf{1}&\textbf{0}\\
 \textbf{1}&\textbf{0}
\end{matrix}
\right)$.

\begin{definition}
Let $\psi\colon \mathcal{M}_n\to
\mathcal{P}^{\mathcal{M}_{n+1}}$ such that:
{\footnotesize
$$
\psi(M)=
\left\{\left(
\begin{array}{c|c|c}
M[1\ldots \lfloor \frac{n}{2} \rfloor ,1\ldots \lfloor \frac{n}{2} \rfloor] &
\begin{matrix}
*\\
*\\
\vdots\\
*\\
\end{matrix} &
M[1\ldots \lfloor \frac{n}{2} \rfloor ,\lfloor \frac{n}{2} \rfloor +1\ldots n]\\
\hline
\begin{matrix}
*&*&\ldots&\ldots&\ldots&*
\end{matrix}&
*&
\begin{matrix}
*&*&*&\ldots&*
\end{matrix}\\

\hline

M[\lfloor \frac{n}{2} \rfloor +1\ldots n,1\ldots \lfloor \frac{n}{2} \rfloor] &

\begin{matrix}
*\\
*\\
\vdots\\
*\\
\end{matrix}

& M[\lfloor \frac{n}{2} \rfloor +1\ldots n,\lfloor \frac{n}{2} \rfloor +1\ldots n]
\end{array}
\right):*\in \Sigma\right\}.
$$
}
\end{definition}

\noindent
The operator $\psi$ inserts in the matrix $M$ a new column and a new row where the
entries can be chosen from the alphabet without restrictions, while the other entries are
inherited from $M$.

\noindent For example, the matrix
$\left(
\begin{matrix}
\textbf{1}&0&\textbf{0}\\
1&0&0\\
 \textbf{1}&1&\textbf{0}
\end{matrix}
\right) \in \psi\left(
\begin{matrix}
\textbf{1}&\textbf{0}\\
 \textbf{1}&\textbf{0}
\end{matrix}
\right).
$

\bigskip

\noindent

In this section we generate the set, denoted by $BBF_n^q$, of all $n \times n$ bibifix-free matrices
over a $q$-ary alphabet $\Sigma=\{0,1,\ldots q-1\}$.
We distinguish two cases depending on the parity of $n\geq 1$.

\begin{itemize}
\item[$\bullet$]
Let $T\in BBF_n^q$, with $n$ even. It is easy to see that
$\psi(T)\subseteq BBF_{n+1}^q$ and
$BBF_{n+1}^q=\{\psi(T)|T\in BBF_n^q\}$. Indeed, if $T_1$ and $T_2$ are two different
bibifix-free matrices, then $\psi(T_1)\cap\psi(T_2)=\emptyset$ and if $T'\in BBF_{n+1}^q$, then there exists
$T\in BBF_n^q$ such that $T'\in\psi(T)$. In other words, the set $\{\psi(T)|T\in BBF_n^q\}$ is a
partition of $BBF_{n+1}^q$.

\item[$\bullet$]
On the other hand, in the case of $n$ odd, it may happen that $\psi(T)$ contains some
matrices which are not bibifix-free. For example,

if $T=\left(
		\begin{matrix}
                    1&0&0&0&0\\
		    0&0&0&0&0\\
		    0&0&1&0&0\\
		    0&0&0&0&0\\
		    0&0&0&0&0\\

		  \end{matrix}
\right)
		  $,
then $T'=\left(
		\begin{matrix}
                    1&0&0&0&0&0\\
		    0&0&0&0&0&0\\
		    0&0&0&0&0&0\\
		    0&0&0&1&0&0\\
		    0&0&0&0&0&0\\
		    0&0&0&0&0&0\\

		  \end{matrix}
		  \right) \in \psi(T)
		  $
		but $T'
		\notin{BBF_{n+1}^q}		
	     $
	     since it contains the bibifix $\left(
	     \begin{matrix}
	                                     1&0&0\\
					     0&0&0\\
					     0&0&0\\
	                                    \end{matrix}
\right)$.

More generally, it is possible to show that in the set $\psi(T)$ the matrices $T'$ which are not bibifix-free
are exclusively the ones having the bibifix of dimension $\frac{n+1}{2}\times \frac{n+1}{2}$ and this bibifix belongs
to $BBF_{\frac{n+1}{2}}^q$.

Formalizing, we get the following proposition.

\begin{proposition}
 If $n$ is odd and $T\in BBF_n^q$, let $T'\in\psi(T)$.
Then, $T'\notin BBF_{n+1}^q$ if and only
if
$T'$ has one and only one bibifix of dimension $\frac{n+1}{2}\times \frac{n+1}{2}$ belonging to
$BBF_{\frac{n+1}{2}}^q.$

\end{proposition}

\emph{Proof.} \quad If $T'[1\ldots\frac{n+1}{2},1\ldots\frac{n+1}{2}]=T'[\frac{n+1}{2}+1\ldots
n,\frac{n+1}{2}+1\ldots n]$, then obviously $T'\notin BBF_{n+1}^q$.

On the other side, let $T'\notin BBF_{n+1}^q$ and suppose ad absurdum that
$T'[1\ldots\frac{n+1}{2},1\ldots\frac{n+1}{2}]\neq T'[\frac{n+1}{2}+1\ldots
n,\frac{n+1}{2}+1\ldots n]$. Then there exists $i$, with $1\leq i\leq \frac{n+1}{2}$, such that
$T'[1\ldots i,1\ldots i]= T'[\frac{n+1}{2}+1+i\ldots
n,\frac{n+1}{2}+1+i\ldots n]$ since a bibifix must occur in $T'$. This bibifix necessarily occurs in $T$
as $T'\in\psi(T)$. This is a contradiction for $T\in BBF_n^q$. Note that this argument
shows also that the dimension of the bibifix can not be less than $\frac{n+1}{2}\times \frac{n+1}{2}$.

Now we have to prove that
$T'[1\ldots\frac{n+1}{2},1\ldots\frac{n+1}{2}]\in BBF_{\frac{n+1}{2}}^q$. For
this purpose, if it is not, then there exists $i$, with $1\leq i\leq \frac{n+1}{4}$,
such that
$T'[1\ldots i,1\ldots i]=T'[\frac{n+1}{4}+1+i\ldots \frac{n+1}{2},\frac{n+1}{4}+1+i\ldots \frac{n+1}{2}]$.
Since $T'\left[1\ldots\frac{n+1}{2},1\ldots\frac{n+1}{2}\right]=T'\left[\frac{n+1}{2}+1\ldots n,\frac{n+1}{2}+1\ldots
n\right]$ (proved in the previous paragraph), $T$ would have a bibifix of dimension $i\times i$, with $i\leq
\frac{n+1}{4}$
against the hypothesis $T\in BBF_n^q$.
\cvd

Note that, Proposition 3.1 describes the matrices of dimension $(n+1)\times (n+1)$
which are not bibifix-free once the operator $\psi$
is applied to all the matrices $T\in BBF_n^q$. More precisely they are the matrices of
the set $\{\varphi(D)|D\in BBF_{\frac{n+1}{2}}^q\}$. The following proposition summarizes the previous
results:

\begin{proposition}
 If $n$ is odd, then
 $$
 BBF_{n+1}^q=\{\psi(T)|T\in BBF_n^q\}
 \setminus\{\varphi(D)|D\in BBF_{\frac{n+1}{2}}^q\}
 $$
\end{proposition}

\end{itemize}

We are now able to give a formula for the cardinality of $BBF_n^q$, denoted by
$|BBF_{n}^q|$.

\begin{proposition}

\begin{equation*}
|BBF_{n}^q|=
\left\{
\begin{array}{ll}
q & \quad\mbox{if}\quad n=1,\\
\\
q^{2n-1}|BBF_{n-1}^q| & \quad\mbox{if}\quad n \ odd,\\
\\
q^{2n-1}|BBF_{n-1}^q| - q^{n^2/2} |BBF_{\frac{n}{2}}^q| &\quad\mbox{if}\quad n
\ even\ .
\end{array}
\right.
\end{equation*}
\end{proposition}

\bigskip
As in the linear case, it is worth to analyze the properties of particular sets of matrices having biprefixes which are
not bisuffixes. Before starting this study, we provide the following two definitions useful in the next paragraph.

\begin{definition}
Two distinct $n \times n$ bibifix-free matrices $T, T'\in BBF_n^q$ are \emph{cross-bibifix-free}
if and only if no biprefix of $T$ is also a bisuffix of $T'$ and viceversa.
\end{definition}

\noindent For example, considering the set $BBF_5^4$, the matrices $T=\left(
  \begin{matrix}
    1&1&1& 1& 3\\
    1 &1 &1 &1& 2\\
    1 &2& 2& 1& 0\\
    0& 3& 1& 1& 0\\
    1& 0& 0& 0& 0
   \end{matrix}
\right)
$ and $T'=\left(
  \begin{matrix}
    1&1&1& 1& 3\\
    1 &1 &1 &0& 2\\
    2 &2& 0& 2& 1\\
    3& 3& 1& 0& 0\\
    2& 0& 0& 0& 2
   \end{matrix}
\right)$ are cross-bibifix-free.
\begin{definition}
A subset of $BBF_{n}^q$ is said to be \emph{cross-bibifix-free set} if
and only if for each distinct $T, T'$ in this set, $T$ and $T'$ are
cross-bibifix-free. This set is said to be \emph{non-expandable} on
$BBF_{n}^q$ if and only if the set obtained by adding any other matrix
is not a cross-bibifix-free set. A non-expandable cross-bibifix-free
set on $BBF_{n}^q$ having maximal cardinality is called
\emph{maximal cross-bibifix-free set} on $BBF_{n}^q$.

\end{definition}

In the next section we are interested in
a possible generation of non-expandable
cross-bibifix-free sets.

\section{On the non-expandability of cross-bibifix-free sets}

Fixed an alphabet $\Sigma=\{0,1,\ldots, q-1\}$, once the bibifix-free set of a given dimension $n\times n$ is
generated, our
aim is the definition of a non-expandable cross-bibifix-free set of square $n\times n$ matrices, denoted by $ CBBF_n^q$.

The constructive method for $ CBBF_n^q$ moves from a non-expandable cross-bifix-free set $A_n^q$ of $q$-ary $n$ length
words. More precisely, for each $u\in A_n^q$, we consider the set of matrices $\mathcal{M}_n(u)$ where
each matrix is obtained by posing $u$ as main diagonal while all the other entries are arbitrarily chosen from
$\Sigma$. Then we define $CBBF_n^q=\{\bigcup \mathcal{M}_n(u)|u\in A_n^q\}$.

From its construction, it is not difficult to realize that the cardinality of $ CBBF_n^q$ is given by
$| CBBF_n^q|=q^{n^2-n}|A_n^q|$. For this reason, it is natural to consider the non-expandable
cross-bifix-free set $A_n^q$ with the largest cardinality in order to obtain the largest cardinality for
$ CBBF_n^q$. To our knowledge the non-expandable cross-bifix-free set with the largest cardinality is given by
the set provided  in \cite{7}, whose definition is also recalled in the rest of this section.
We will prove that, considering such a set of words, then $CBBF_n^q$ is a non-expandable cross-bibifix-free
set of matrices.

For the sake of simplicity, first we analyze the case of a binary alphabet, then we generalize the results to
a $q$-ary alphabet $\Sigma=\{0,1,\ldots,q-1\}$.

\subsection{The binary case}
In this section we provide a non-expandable cross-bibifix-free set $ CBBF _n^2$ of square matrices of fixed
dimension $n\times n$, with $n\geq2$, in the binary case ($ CBBF _n^2\subset BBF_{n}^2$).

\noindent First, we note that the set $BBF_{n}^2$ can be partitioned
in $BBF_{n}^2=\mathcal{U}_n^2\cup\mathcal D_n^2$ where
$\mathcal U_n^2$ contains the square matrices $U$ of dimension $n\times n$  such that $U[1,1]=1$ and $U[n,n]=0$ and
$\mathcal D_n^2$ contains the square matrices $D$ of dimension $n\times n$  such that $D[1,1]=0$ and $D[n,n]=1$.
Clearly, each cross-bibifix-free set is completely contained either in $\mathcal D_n^2$ or in $\mathcal U_n^2$, since
$U[1,1]=D[n,n]=1$ and $D[1,1]=U[n,n]=0$ for any $U \in \mathcal U_n^2$ and $D \in \mathcal D_n^2$.
The set $ CBBF_n^2$ we are going to construct is contained in $\mathcal{U}_n^2$, for $n\geq2$.

Analogously, the set $BF_n^2$ of all bifix-free binary strings can be partitioned in $BF_n^2=\mathcal{L}_n^2 \cup
\mathcal{R}_n^2$, where $\mathcal{L}_n^2$ contains the strings $u$ of length $n$ such that $u[1]=1$ and $u[n]=0$ and
$\mathcal{R}_n^2$ contains the strings $v$ of length $n$ such that $v[1]=0$ and $v[n]=1$.

\bigskip

We describe now
the considered non-expandable cross-bifix-free set of words, denoted by
$S_{n,2}^{(k)}$, in order to generate $ CBBF _n^2$. The set $S_{n,2}^{(k)}$ is formed by length $n$ words over
the binary
alphabet containing a particular sub-word avoiding $k$
consecutive $1$s.

In the sequel we briefly summarize its definition, nevertheless for more details
about its features we refer the reader to \cite{7}. With respect to the original definition here we replace the $0$s with $1$s.

Let $n \geq 3$ and $1 \leq k \le n-2$. The non-expandable cross-bifix-free set $S_{n,2}^{(k)}$
is the set of all length $n$ words $s[1] s[2] \cdots s[n]$ satisfying:
\begin{itemize}
\item $s[1] = \dots = s[k] = 1$;
\item $s[k+1] = 0$; $s[n] = 0$;
\item the sub-word $s[k+2] \dots s[n-1]$ does not contain $k$ consecutive $1$s.
\end{itemize}

Note that, for any fixed $n$, the cardinality of $S_{n,2}^{(k)}$ depends on $k$. In the rest of this paragraph we
assume that the value of $k$ is the one giving the maximum cardinality (for more details see \cite{7}) and this set is
denoted by $S_n^2$.

\begin{proposition}\label{no-expa}

Suppose $S_n^2=\{w_1,w_2,\ldots,w_{|S_n^2|}\}$.

\noindent Denoting $w_i=w_i[1]w_i[2]
w_i[3]\ldots w_i[n]$, $i=1,2,\ldots,|S_n^2|$,
the set $ CBBF_n^2\subset\mathcal{U}_n^2$ given by
$$
 CBBF_n^2=\left\{\left(
\begin{matrix}
 w_i[1] & * & * &  \dots  & * \\
 * & w_i[2] & * &  \dots  & * \\
 * & * & w_i[3] &  \dots & *\\
    \vdots & \vdots & \vdots & \ddots & \vdots \\
    * & \dots & * & *  & w_i[n]
\end{matrix}
\right)
: *\in\{0,1\},\forall i
\right\}
$$
\noindent is a non-expandable cross-bibifix-free set on $BBF_{n}^2$, with $n\geq 3$.

\end{proposition}

\emph{Proof.} \quad First of all, we prove that $ CBBF_n^2$ is a cross-bibifix-free set, for any fixed $n \geq 3$.

Let $C, C' \in  CBBF_n^2$ having $w_i$ and $w_j$, possibly the same, as their main diagonal, respectively,
with $C\neq C'$.
Each biprefix $C[1 \ldots r, 1 \ldots r]$ of $C$, with $r\leq n$, is different from any bisuffix
$C'[n-r+1 \ldots n, n-r+1 \ldots n]$ of $C'$ for any entries
$*\in\{0,1\}$,  since $w_i[1]\ldots w_i[r] \neq w_j[n-r+1]\ldots w_j[n]$ for each $1 \leq i,j \leq |S_n^2|$,
being $S_n^2$ cross-bifix-free set. Then $ CBBF_n^2$ is a
cross-bibifix-free set, and
$ CBBF_n^2\subset\mathcal{U}_n^2$.
\bigskip

As far as the non-expandability of $ CBBF_n^2$ is concerned, it can be first observed that, by using a similar argument, the set is
not expandable by matrices of the form
$$
B=\left(
\begin{matrix}
 b_i[1] & * & * &  \dots  & * \\
 * & b_i[2] & * &  \dots  & * \\
 * & * & b_i[3] &  \dots & *\\
    \vdots & \vdots & \vdots & \ddots & \vdots \\
    * & \dots & * & *  & b_i[n]
\end{matrix}
\right)
$$
where $B\in BBF_n^2$, $b_i=b_i[1]b_i[2]\ldots b_i[n]$ is a word of $BF_n^2$ but
$b_i\notin S_n^2$: indeed,
since $S_n^2$ is non-expandable, each prefix (suffix) of $b_i$ is a suffix (prefix) of $w_j$, for some $j$,
then each biprefix (bisuffix) of $B$, for any choice of the entries not belonging to
the main diagonal, is the bisuffix (biprefix) of some matrix in $ CBBF_n^2$, for any fixed $n \geq 3$.

We observe that the particular
choice of the non-expandable
cross-bifix-free set we have considered does not affect the proof up to this point. From now, on the contrary, the
particular structure of $S_n^2$ is crucial.

\bigskip

We now investigate on the possibility to expand $ CBBF_n^2$ with matrices $M\in BBF_n^2$ but where
the main diagonal $m_i\notin BF_n^2$. In other words, $m_i$ presents a bifix $1\alpha0$ of length less or
equal to $\lfloor\frac{n}{2}\rfloor$. Then $m_i=1\alpha0\varphi 1\alpha0$, where
$|1\alpha0|\leq\lfloor\frac{n}{2}\rfloor$ and $\varphi$, $\alpha$ are binary strings of suitable length, possibly
empty, so that $n\geq 4$.

\noindent
The matrix
$
M=\left(
\begin{matrix}
\bold{1}&0&0&0&0&0\\
0&\bold{1}&0&0&0&0\\
0&0&\bold{0}&0&0&0\\
0&0&0&\bold{1}&0&0\\
0&0&0&1&\bold{1}&0\\
0&0&0&1&1&\bold{0}\\
\end{matrix}
\right)
$
is an example of a bibifix-free matrix where the main diagonal is not bifix-free.

We can show that for each $m_i \notin BF_n^2$, there exists a string $w_i=1^k0 \gamma 0 \in
S_n^2$
having a prefix or a suffix of suitable length equal to a suffix or a prefix (of the same length) of $m_i$. Clearly, we
consider only those words $m_i \notin BF_n^2$ beginning with 1. Such a word can be factorized as
$m_i=1\alpha0\varphi 1\alpha0$, where $|1\alpha0| \leq \lfloor \frac{n}{2} \rfloor$ with $\alpha$ and
$\varphi$ possibly empty. We can distinguish two cases:
\begin{itemize}
 \item [A)] The bifix $1\alpha0$ contains at least $k$ consecutive 1s. In this case considering the rightmost
sequence $1^k$ the bifix can be written as $1 \alpha 0= \beta 1^k 0 \beta'$ where $\beta'$ does not contain $1^k$ and
$\beta$ and $\beta'$ may be empty. It is easily seen that the set $S_n^2$ contains, for example, the word
$1^k0\beta'0^{n-k-1-|\beta'|}$ which presents the prefix $1^k0\beta'$ equal to the suffix of $m_i$.
Note that in this case, being $k\geq1$, the bifix $1\alpha0$ with the smallest length is $10$, then the length $n$ of
$m_i$ is greater than or equal to 4.

\item [B)] The bifix $1 \alpha 0$ does not contain $k$ consecutive 1s. Then $1 \alpha 0= 1^m0 \beta$ with $m<k$
and $\beta$ possibly empty. In this case the prefix $1^m0$ of $m_i$ occurs as a suffix in $1^k0^{n-k-m-1}1^m0 \in
S_n^2$.
Note that at least one zero must occur between $1^k$ and $1^m$, then $n-k-m-1\geq1$. Moreover, since in this case
$k\geq2$ and $m\geq1$, we have $n\geq5$.

\end{itemize}

Obviously, for $n=3$, there do not exist bibifix-matrices where the main diagonal contains a bifix. Then, summarizing,
moving from $S_n^2$, for $n\geq3$, the set $ CBBF_n^2$ provides a non-expandable
cross-bibifix free set on $BBF_n^2$.

\cvd

For the sake of completeness, in the case $n=4$, it is $|S_4^2|=1$ and we can assume both
$S_4^2=\{1000\}$ or $S_4^2=\{1100\}$ considering $k=1$ or $k=2$, respectively.
Assuming $S_4^2=\{1100\}$, the cross-bibifix-free set $ CBBF_4^2$, according to the
definition given in Proposition 4.1, would be equal to
$
\left\{\left(
\begin{matrix}
 1 & * & * & *\\
 * & 1 & * & *\\
 * & * & 0 & *\\
 * & * & * & 0
\end{matrix}
\right)
: *\in\{0,1\}
\right\}
$. We can note that a such cross-bibifix-free set can be expanded, for example, with the matrix
$M=\left(\begin{matrix}
 1 & 0 & 0 & 0\\
 0 & 0 & 0 & 0\\
 0 & 0 & 1 & 1\\
 0 & 0 & 1 & 0
\end{matrix}
\right) \in BBF_2^4$.
In order to obtain a non-expandable cross-bibifix-free set, we have to consider $S_4^2=\{1000\}$ as
the main diagonal of the matrices. Then, we define
$$ CBBF_4^2=
\left\{\left(
\begin{matrix}
 1 & * & * & *\\
 * & 0 & * & *\\
 * & * & 0 & *\\
 * & * & * & 0
\end{matrix}
\right)
: *\in\{0,1\}
\right\}
\ ,$$
which is easily seen to be a non-expandable cross-bibifix free set, following a similar argument used in in the proof
of Proposition 4.1.

\bigskip

\noindent {\bf{Remark.}}
Really, moving from any cross-bifix-free set of words, it is always possible to obtain a cross-bibifix-free set of
matrices using the technique outlined in Proposition 4.1, regardless of the non-expandability.

\subsection{The $q$-ary case}

The definition of the cross-bifix-free set $S_{n,q}^{(k)}$ of length $n$ words $s[1]s[2]\ldots s[n]$, with $n\geq 3$
and $1 \leq k \leq n-2$ is given by \cite{7}:

\begin{itemize}
\item $s[1] = \dots = s[k] = 1$;
\item $s[k+1] \neq 1$; $s[n] \neq 1$;
\item the sub-word $s[k+2] \dots s[n-1]$ does not contain $k$ consecutive $1$s.
\end{itemize}

As in the previous case, here we assume that the value of $k$ is the one giving the maximum cardinality, once $n$ is
fixed. We denote this set with $S_n^q$.

\begin{proposition}\label{no-expa}
Suppose $S_n^q=\{w_1,w_2,\ldots,w_{|S_n^q|}\}$.

\noindent Denoting $w_i=w_i[1]w_i[2]
w_i[3]\ldots w_i[n]$, $i=1,2,\ldots,|S_n^q|$,
the set $ CBBF_n^q$ given by
$$
 CBBF_n^q=\left\{\left(
\begin{matrix}
 w_i[1] & * & * &  \dots  & * \\
 * & w_i[2] & * &  \dots  & * \\
 * & * & w_i[3] &  \dots & *\\
    \vdots & \vdots & \vdots & \ddots & \vdots \\
    * & \dots & * & *  & w_i[n]
\end{matrix}
\right)
: *\in\Sigma, \forall i
\right\}
$$
  \noindent is a non-expandable cross-bibifix-free set on $BBF_{n}^q$, with $n\geq 3$.
\end{proposition}

\emph{Proof.} \quad First, we prove that $ CBBF_n^q$ is a cross-bibifix-free set, for any fixed $n \geq 3$.
Let $C, C' \in  CBBF_n^q$ having $w_i$ and $w_j$, possibly the same, as their main diagonal, respectively,
with $C\neq C'$.
Each biprefix $C[1 \ldots r, 1 \ldots r]$ of $C$ (with $r\leq n$) is different from any bisuffix
$C'[n-r+1 \ldots n, n-r+1 \ldots n]$ of $C'$ for any entries
$*\in \Sigma$,  since $w_i[1]\ldots w_i[r] \neq w_j[n-r+1]\ldots w_j[n]$ for each $1 \leq i,j \leq |S_n^q|$,
being $S_n^q$ cross-bifix-free set. Then $ CBBF_n^q$ is a
cross-bibifix-free set.
\bigskip

For the non-expandability of $ CBBF_n^q$ we have to prove that for any matrix
$M \in BBF_n^q \backslash  CBBF_n^q$ there exits a matrix in $ CBBF_n^q$ having its biprefix
(bisuffix) equal to a bisuffix (biprefix) of $M$. As each matrix $C \in  CBBF_n^q$ admits $C[1,1]=1$ and
$C[n,n] \neq 1$ then we can only consider the matrices $M$ having $M[1,1]=1$ and $M[n,n] \neq 1$, otherwise we can
easily note that $C[1,1]=M[n,n]=1$ for any $C \in  CBBF_n^q$, and for each $s \in \Sigma \backslash \{1\}$
there exists a matrix $C \in  CBBF_n^q$ such that  $C[n,n]=M[1,1]=s \neq 1$.

The set $ CBBF_n^q$ is not expandable with matrices $M \in BBF_n^q$ having as their main
diagonal $m_i$ a word of $BF_n^q$. Indeed, since $S_n^q$ is non-expandable, there is a prefix
(suffix) of $m_i$ equal to a suffix (prefix) of $w_i \in S_n^q$, for some $i$, then there exists a biprefix
(bisuffix) of $M$, for any choice of the entries not belonging to the main diagonal, equal to the bisuffix (biprefix) of
some matrix in $ CBBF_n^q$.

\bigskip

We now investigate on the possibility to expand $ CBBF_n^q$ with matrices $M\in BBF_n^q$ having
the main diagonal $m_i\notin BF_n^q$. In particular, the main diagonal $m_i$ of $M$ presents a bifix $1\alpha
d$ of length less or equal to $\lfloor\frac{n}{2}\rfloor$, with $d \in \Sigma \backslash \{1\}$. So, we can consider
$m_i=1\alpha d\varphi 1\alpha d$, where $|1\alpha d|\leq\lfloor\frac{n}{2}\rfloor$ and $\alpha, \varphi$ are two
$q$-ary strings of suitable length, possibly empty, so that $n\geq 4$.

\medskip

We can show that for each $m_i \notin BF_n^q$, there exists a string $w_i \in
S_n^q$ having a prefix or a suffix of suitable length equal to a suffix or a prefix (of the same length) of
$m_i$.
We can distinguish two cases:
\begin{itemize}
 \item [A)] The bifix $1\alpha d$ contains at least $k$ consecutive 1s. In this case considering the rightmost
sequence $1^k$ the bifix can be written as $1 \alpha d= \beta 1^k l \beta'$, with $l \neq 1$, where $\beta'$ does
not contain $1^k$ and $\beta$ and $\beta'$ may be empty (if $\beta'$ is empty then $l$ coincides with $d$). It is
easily seen that the set $S_n^q$ contains, for example, the word
$1^k l\beta'0^{n-k-1-|\beta'|}$ which presents the prefix $1^k l\beta'$ equal to the suffix of $m_i$.
Note that in this case, being $k\geq1$, the bifix $1\alpha d$ with the smallest length is $1 d$, then the length $n$ of
$m_i$ is greater than or equal to 4.

\item [B)] The bifix $1 \alpha d$ does not contain $k$ consecutive 1's. Then $1 \alpha d= 1^m l \beta$, with $m<k$ and
$l \neq 1$, and $\beta$ possibly empty (if $\beta$ is empty then $l$ coincides with $d$). In this case the prefix $1^m
l$ of $m_i$ occurs as a suffix in $1^k0^{n-k-m-1}1^m l \in S_n^q$. Note that at least one symbol different
from 1 must occur between $1^k$ and $1^m$, then $n-k-m-1\geq1$. Moreover, since in this case $k\geq2$ and $m\geq1$, we
have $n\geq5$.

\end{itemize}

Obviously, for $n=3$, there do not exist bibifix-matrices where the main diagonal contains a bifix. Then, summarizing,
moving from $S_n^q$, for $n\geq3$ the set $ CBBF_n^q$ provides a non-expandable cross-bibifix free
set on $BBF_n^q$.

\cvd

\section{A Gray code for $CBBF_n^q$}
Once a class of objects is defined, in our case matrices, often it could be useful to list or generate
them according to a particular criterion. A special way to do this is their generation
in a way such that any two consecutive matrices differ as little as
possible: i.e. Gray codes \cite{11}. In our case we are going to provide a Gray code, denoted by
$\mathcal{CBBF}_n^q$, for the set $CBBF_n^q$ where two consecutive matrices differ only in one entry.

In the following we will use the notations below:

\begin{itemize}
\item For a list of words $\mathcal L$, $\overline{\mathcal L}$ denotes the list obtained by covering
      $\mathcal L$ in reverse order; and for $i\geq 0$, $(\mathcal L)^{\underline i}$ denotes the list $\mathcal
L$ if $i$ is even, and the list $\overline{\mathcal L}$ if $i$ is odd;

\item If $\alpha$ is a word, then $\alpha\cdot \mathcal L$ is the list obtained by concatenating $\alpha$
      to each word of $\mathcal L$;

\item For two lists $\mathcal L$ and $\mathcal L'$, $\mathcal L\circ \mathcal L'$
      denotes their concatenation, and
      for two integers, $p\leq r$, and the lists
      $\mathcal L_p,\mathcal L_{p+1},\ldots,\mathcal L_r$,
      we denote by $\concc{i=p}{r}\mathcal L_i$ the list
      $\mathcal L_p\circ\mathcal L_{p+1}\circ\ldots\circ \mathcal L_r$;

\end{itemize}

In order to obtain the Gray code $\mathcal{CBBF}_n^q$ we are going to arrange together two well-known Gray codes in literature.

The first one is presented in \cite{3} and it is a Gray code, denoted by
$\mathcal{S}_n^q$, for the cross-bifix-free set of words $S_n^q$.
In this section, we use it for the main diagonals of the matrices in $\mathcal{CBBF}_n^q$.
For example, the Gray code $\mathcal{S}_3^3=\{100,102,122,120\}$ is applied to the main diagonals of the matrices in $\mathcal{CBBF}_3^3$.

The second one is a Gray code list for the set of words of a certain length over the
$q$-ary alphabet $\Sigma=\{0,1, \ldots, q-1 \}$. Here, we use it for the entries (not belonging to the
main diagonal) of the matrices in $\mathcal{CBBF}_n^q$.
Such a Gray code is an obvious generalization of the Binary Reflected Gray Code
\cite{11} to the $q$-ary alphabet and it is the list
$\mathcal{G}_{n,q}$ for the set of the length $n$ words over $\Sigma$
defined in \cite{10,14} where it is also
shown that $\mathcal{G}_{n,q}$ is a Gray code with Hamming distance~$1$.
We recall that the Hamming distance between two successive words
in a Gray code list is the number of positions where the two words differ.
The list $\mathcal{G}_{n,q}$ is defined as:

\begin{equation*}
\label{g_nq}
\mathcal{G}_{n,q}=\left\{
\begin{array}{cr}
\epsilon &\ \mathrm{if}\ n=0,\\
\\
\concc{i=0}{q-1}\,i\cdot (\mathcal{G}_{n-1,q})^{\underline i} & \mathrm{if}\ n>0,
\end{array}
\right.
\end{equation*}
where $\epsilon$ is the empty word.
%
 For example, considering $\Sigma=\{0,1,2\}$ and $n=6$, then we have:
$$
\begin{array}{rcl}
\mathcal{G}_{6,3}&=&\{000000,000001,000002, \ldots, 022221,022222,122222,122221, \ldots \\
&& \ldots, 100002,100001,100000,200000,200001, \ldots, 222221,222222\}
\end{array}
$$

Since the entries not belonging to the main diagonal of a given matrix can be arranged in a linear word (see details
below), the list $\mathcal{G}_{n^2-n,q}$ provide a Gray code for the extra-diagonal entries of the matrices in
$\mathcal{CBBF}_n^q$.
So, a crucial step of our strategy consists in the linearization of the elements of the matrix (regardless of the
ones belonging to the main diagonal) in order to have a mono-dimensional structure. There are several methods to do
this, in the following we describe the one we adopt: given a matrix $C$, each entry $C[i,j]$ (avoiding the elements of
the diagonal) is associated to $w[k]$ ranging $k$ from 1 up to $n^2-n$, reading the entries from top to bottom and from
left to right, starting from $C[2,1]$ (associated to $w[1]$) and first completing the lower triangular sub-matrix up to
$C[n,n-1]$ associated to $w[(n^2-n)/2]$, and then considering the upper triangular sub-matrix starting from $C[1,2]$
corresponding to $w[(n^2-n)/2+1]$ up to $C[n-1,n]$ corresponding to $w[n^2-n]$. A clarifying example illustrates our
procedure: the extra-diagonal elements of the $4\times 4$ matrix
$C=\left(
\begin{matrix}
 C[1,1] & C[1,2] & C[1,3] & C[1,4] \\
 C[2,1] & C[2,2] & C[2,3] & C[2,4] \\
 C[3,1] & C[3,2] & C[3,3] & C[3,4] \\
 C[4,1] & C[4,2] & C[4,3] & C[4,4]
\end{matrix}
\right)$
are marked with the indexes
$\left(
\begin{matrix}
   *  & w[7] & w[8] & w[10] \\
 w[1] &   *  & w[9] & w[11] \\
 w[2] & w[4] &   *  & w[12] \\
 w[3] & w[5] & w[6] &   *
\end{matrix}
\right)$
and form the linear word $w[1]w[2]\ldots w[12]$.

\bigskip

Formalizing, the elements of the matrix $C$ not belonging to the main diagonal form a word
$w=w[1]w[2]\ldots w[n^2-n]$ such that $w[f_{i,j}]$ corresponds to $C[i,j]$ (with $i\neq j$) where:

\begin{equation*}
\label{g_nq}
f_{i,j}=\left\{
\begin{array}{cr}
n(j-1)+i-\frac{j(j-1)}{2} &\ \mathrm{if}\ i>j\ ,\\
\\
\frac{n(n-1)}{2}+\frac{j(j-1)}{2}+i-j+1 & \mathrm{if}\ i<j\ .
\end{array}
\right.
\end{equation*}

\noindent
On the other side, given a word $w=w[1]w[2]\ldots w[n^2-n]$ it is possible to construct the matrix $C$, regardless of
the main diagonal, by setting $C[i,j] =w[f_{i,j}]$ ($i\neq j$). For the sake of clearness, in the case $q=3$
(so that $\Sigma=\{0,1,2\}$) and $n=4$, if $w=121201100020$, then
$C=\left(
\begin{matrix}
 * & 1 & 0 & 0 \\
 1 & * & 0 & 2 \\
 2 & 2 & * & 0 \\
 1 & 0 & 1 & *
\end{matrix}
\right)$.

\bigskip
At this point, we are able to describe the construction of the Gray code $\mathcal{CBBF}_n^q$.
Let $\mathcal{S}_n^q=\{s_1,s_2,\ldots,s_{|\mathcal{S}_n^q|}\}$ and
$\mathcal{G}_{n^2-n,q}=\{g_1,g_2,\ldots,g_{q^{n^2-n}}\}$. First we define the list
$\left(s_k,\mathcal{G}_{n^2-n,q}\right)$ of $q^{n^2-n}$ matrices having
all the same diagonal $s_k$  while the other entries are obtained by $g_1, g_2$ up to $g_{q^{n^2-n}}$ as in the
previous example. The reader can easily see that $\left(s_k,\mathcal{G}_{n^2-n,q}\right)$ is a Gray code for each
$s_k \in \mathcal{S}_n^q$, with $k=1,\ldots, |\mathcal{S}_n^q|$.
Then, the Gray code $\mathcal{CBBF}_n^q$ for the set $CBBF_n^q$ can be defined as:
$$
\mathcal{CBBF}_n^q= \concc{k=0}{|\mathcal{S}_n^q|-1}\,\left(s_{k+1}, (\mathcal{G}_{n^2-n,q})^{\underline k}\right).
$$

For example, $\mathcal{CBBF}_3^3$ is given by

$$
\begin{array}{l}
\left(
\begin{matrix}
 1 & 0 & 0 \\
 0 & 0 & 0 \\
 0 & 0 & 0
\end{matrix}
\right),
\left(
\begin{matrix}
 1 & 0 & 0 \\
 0 & 0 & 1 \\
 0 & 0 & 0
\end{matrix}
\right),
\ldots,
\left(
\begin{matrix}
 1 & 2 & 2 \\
 2 & 0 & 1 \\
 2 & 2 & 0
\end{matrix}
\right),
\left(
\begin{matrix}
 1 & 2 & 2 \\
 2 & 0 & 2 \\
 2 & 2 & 0
\end{matrix}
\right),
\\\\
\left(
\begin{matrix}
 1 & 2 & 2 \\
 2 & 0 & 2 \\
 2 & 2 & 2
\end{matrix}
\right),
\left(
\begin{matrix}
 1 & 2 & 2 \\
 2 & 0 & 1 \\
 2 & 2 & 2
\end{matrix}
\right),
\ldots,
\left(
\begin{matrix}
 1 & 0 & 0 \\
 0 & 0 & 1 \\
 0 & 0 & 2
\end{matrix}
\right),
\left(
\begin{matrix}
 1 & 0 & 0 \\
 0 & 0 & 0 \\
 0 & 0 & 2
\end{matrix}
\right),
\\\\
\left(
\begin{matrix}
 1 & 0 & 0 \\
 0 & 2 & 0 \\
 0 & 0 & 2
\end{matrix}
\right),
\left(
\begin{matrix}
 1 & 0 & 0 \\
 0 & 2 & 1 \\
 0 & 0 & 2
\end{matrix}
\right),
\ldots,
\left(
\begin{matrix}
 1 & 2 & 2 \\
 2 & 2 & 1 \\
 2 & 2 & 2
\end{matrix}
\right),
\left(
\begin{matrix}
 1 & 2 & 2 \\
 2 & 2 & 2 \\
 2 & 2 & 2
\end{matrix}
\right),
\\\\
\left(
\begin{matrix}
 1 & 2 & 2 \\
 2 & 2 & 2 \\
 2 & 2 & 0
\end{matrix}
\right),
\left(
\begin{matrix}
 1 & 2 & 2 \\
 2 & 2 & 1 \\
 2 & 2 & 0
\end{matrix}
\right),
\ldots,
\left(
\begin{matrix}
 1 & 0 & 0 \\
 0 & 2 & 1 \\
 0 & 0 & 0
\end{matrix}
\right),
\left(
\begin{matrix}
 1 & 0 & 0 \\
 0 & 2 & 0 \\
 0 & 0 & 0
\end{matrix}
\right).
\end{array}
$$

\section{Conclusion and further developments}

The structures we have considered in our paper are exclusively square matrices. A first further improvement of our
study should take into consideration matrices which are not square.
In order to consider a cross-bibifix-free set of $n \times m$ matrices, with $n < m$, a direct extension of our
approach can be easily carried on, in the sense that, given two matrices $C$ and $C'$, they are said
\emph{cross-bibifix-free} if any biprefix $C[1\ldots r,1\ldots r]$ of $C$, with $r<n<m$, is different from any
bisuffix $C[n-r+1\ldots n,m-r+1\ldots m]$ of $C'$ of the same dimension, and viceversa. In other words, we are going to
consider square bibifixes, as in the case of square matrices.

With a similar argument used in Section 4, we consider a cross-bifix-free set $A_n^q$ of $q$-ary $n$ length words, then
we construct the set $CBBF_{n,m}^q$ of $n\times m$ matrices $C$ by posing any two words, also identical, of $A_n^q$ as
the main diagonal of the biprefix $C[1\ldots n,1\ldots n]$ and the bisuffix $C[1\ldots n,m-n+1\ldots m]$, while all
the other entries are symbols of a $q$-ary alphabet $\Sigma$. It is easily seen that $CBBF_{n,m}^q$ is a
cross-bibifix-free set according to the above definition.
Formalizing, given $A_n^q=\{w_1,w_2,\ldots,w_{|A_n^q|}\}$ and $w_i,w_j \in A_n^q$, $i,j=1,2,\ldots, |A_n^q|$, the
cross-bibifix-free set $CBBF_{n,m}^q$ of matrices $C$ is

{\small
$$
\left\{\left(
\begin{matrix}
w_i[1] & * &  \dots  & * & \dots & * & w_j[1] & * & \dots & *\\
* & w_i[2] &  \dots  & * & \dots & * & * & w_j[2] &   \dots  & *\\
\vdots & \ddots & \ddots & \vdots & \dots & \vdots & \vdots & \ddots & \ddots & \vdots\\
* & \dots  & *  & w_i[n] & \dots & * & * & \dots & * & w_j[n]
\end{matrix}
\right)
: * \in \Sigma,\forall i,j
\right\}
$$
}

\noindent where $w_i[k]=C[k,k]$ and $w_j[k]=C[k,m-n+k]$ for each $k=1,2, \ldots, n$.

\bigskip

Another interesting problem to analyze could be the study of set of matrices \emph{unbordered}. In particular,
instead of the bibifixes we have introduced in this work, the concept of \emph{border} could be considered. Following
\cite{9}, a \emph{border} is a $r\times r$ submatrix $P$ of a $n\times n$ matrix $C$ if $P$ occurs in position $[1,1]$,
$[n-r+1,1]$, $[1,n-r+1]$ and $[n-r+1,n-r+1]$. A matrix $C$ is said \emph{unbordered} if $C$ does not present any
$r\times r$ border, for $r=1,2,\ldots, n-1$. It is worth to investigate if all the results obtained in this paper can
be adapted to the unbordered matrices.

\section{Acknowledgements}

This work has been partially supported by the PRIN project ``Automi e linguaggi formali: aspetti matematici ed applicativi'' and GNCS project ``Strutture discrete con vincoli''.

\end{document}